\documentstyle[11pt,newpasp,twoside,epsf]{article}
\begin{document}
\title{Measuring Black Hole Masses Using Ionized Gas Kinematics}

\author{A. J. Barth$^1$, M. Sarzi, L. C. Ho, H.-W. Rix, J. C. Shields, \\
A. V. Filippenko, G. Rudnick, \& W. L. W. Sargent}
\affil{$^1$Harvard-Smithsonian Center for Astrophysics, 60 Garden St.,
Cambridge, MA 01238}

\begin{abstract}
We describe techniques for measuring the central masses of galaxies
using emission-line kinematics observed with the \emph{Hubble Space
Telescope}.  For accurate results, it is necessary to model various
instrumental effects, particularly the blurring due to the telescope
PSF and the width of the spectroscopic aperture.  Observations of
nuclear gas disks often reveal substantial internal velocity
dispersions in the gas, suggesting that the disks may be partially
pressure-supported.  We also describe a technique for fitting
2-dimensional spectroscopic data directly in pixel space.  This method
may be useful for objects such as M84 that show highly complex and
asymmetric line profiles.
\end{abstract}
\section{Introduction}
Supermassive black holes are thought to occur in the nuclei of all
massive galaxies, or at least in those galaxies having a bulge
component.  Recent studies have shown that the black hole mass is
correlated loosely with the host galaxy bulge luminosity (Kormendy \&
Richstone 1995; Magorrian et al.\ 1998) and correlated tightly with
the stellar velocity dispersion of the bulge (Ferrarese \& Merritt
2000; Gebhardt et al.\ 2000).  However, the slope of the
$M_{\mathrm{BH}}-\sigma$ relation, and the amount of intrinsic scatter
in the correlation, remain somewhat controversial.  For further
progress, more black hole masses must be measured, and with the
highest accuracy possible.
 
Disks of ionized gas and dust with radii of tens to hundreds of
parsecs are found in the nuclei of $\sim10-20\%$ of nearby galaxies.
With \emph{HST}\ spectra, it is possible to map out the velocity
structure of these disks and determine the central masses.  In
principle, the measurement is straightforward: the goal is simply to
measure the rotation speed of a circular disk.  However, the region
with the greatest diagnostic power for determining the central mass is
generally at or near the resolution limit of \emph{HST}.  Near the
nucleus, there are steep gradients in rotation velocity and
emission-line surface brightness across the spectroscopic aperture.
In addition, the gas often has a substantial internal velocity
dispersion, which may affect the dynamical properties of the disks.
All of these effects must be taken into account in order to derive
accurate black hole masses from the data.
 
\section{Measurement Techniques}

\textbf{Instrumental Effects.}  The velocity fields in circumnuclear
disks can be measured with the STIS spectrograph on \emph{HST},
usually with a slit width of 0\farcs1 or 0\farcs2.  Due to the
telescope PSF and the nonzero aperture size, the innermost Keplerian
rotation curve is blurred into a smooth velocity gradient across the
nucleus. Only for the largest black holes will the Keplerian portion
of the velocity field be partly resolved.  Kinematic analyses which do
not take into account this instrumental blurring (e.g., Ferrarese et
al.\ 1996) will tend to underestimate the true black hole masses,
because an unblurred model will need a smaller central mass to reach a
given rotation velocity near the nucleus.

We have developed modeling software that performs a complete
simulation of the STIS observation, so that kinematic models for the
rotating disk can be compared with the data in detail.  The code
calculates the propagation of full line profiles through the
spectrograph slit, accounting for PSF blurring, aperture size, and the
apparent wavelength shift of light that enters the slit off-center.
These techniques have been applied to measure the central mass of the
S0 galaxy NGC 3245 (Barth et al.\ 2001).

\noindent\textbf{Emission-line Surface Brightness.}  Observations of
the velocity field are weighted by the surface brightness of emission
lines, which sometimes show a sharp, nearly unresolved ``spike'' at
the nucleus.  The measured velocities near the nucleus can be very
sensitive to the brightness profile of this central spike.  In
addition, a nonuniform distribution of emission-line surface
brightness can distort the shape of the radial velocity curves if a
patch of emission is located off-center in the slit (Figure 1).

\noindent\textbf{Disk Orientation Parameters.}  At least 3 slit
positions are needed to fully constrain the orientation of the disk.
For NGC 3245, we used 5 parallel positions covering the inner
arcsecond.  The off-nuclear slit positions give enough kinematic
information to tightly constrain the disk inclination and the major
axis position angle, eliminating what would otherwise be a significant
source of uncertainty in the analysis.

\noindent\textbf{Intrinsic Velocity Dispersion and Asymmetric Drift.}
Our modeling code includes all of the major sources of line broadening
that contribute to the observed linewidths.  These include the
point-source line-spread function, the broadening due to the nonzero
width of the slit, the diffusion of charge on the CCD, and the
rotational broadening for the portion of the disk subtended by the
slit aperture.  The models for NGC 3245 demonstrate that the disk
cannot be dynamically cold: the gas has an intrinsic velocity
dispersion which rises to $\sim150$ km s$^{-1}$ at the nucleus.

Substantial internal velocity dispersions are observed in most
circumnuclear disks of this type.  This suggests that the disks may be
partially pressure-supported, and that the black hole mass inferred
from pure circular rotation models may be an underestimate.  For NGC
3245, we apply a correction for this effect using the asymmetric drift
equation of stellar dynamics; in this case the effect on the black
hole mass is only 12\%, but it may be larger for other galaxies with
more turbulent disks.

For NGC 3245, we find $M_{\mathrm{BH}} = (2.1 \pm 0.5) \times10^8$
$M_\odot$.  The error budget includes uncertainties due to the
kinematic model fitting, the measurement of the stellar luminosity
profile, the distance to the galaxy, and differences between the
kinematics of H$\alpha$ and [N II].

\begin{figure}
\plotone{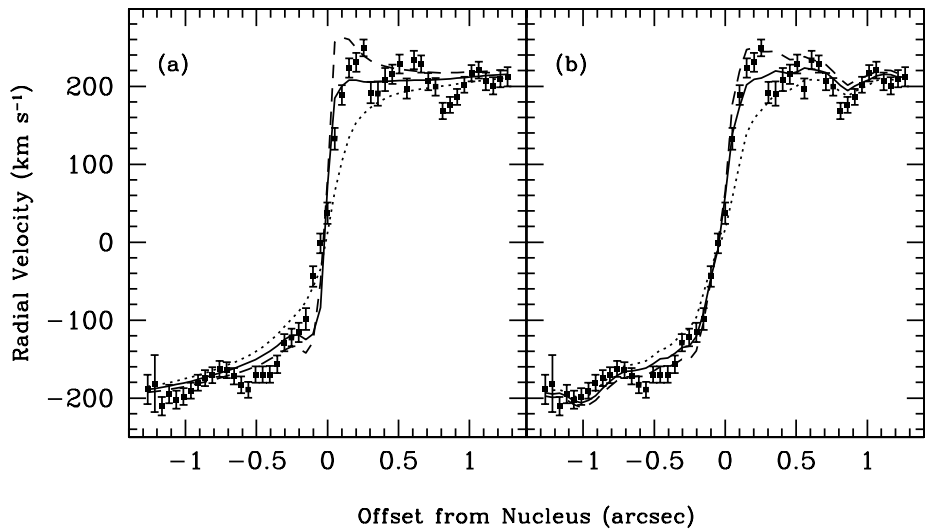}
\caption{Central velocity curve for NGC 3245 measured from STIS data,
using the 0\farcs2-wide slit.  The model curves are calculated for
$M_{\mathrm{BH}}=0, 2\times10^8$, and $4\times10^8$ $M_\odot$ and a
fixed value of the stellar mass-to-light ratio.  (\emph{a}) Model
curves calculated assuming a smooth emission-line surface brightness
distribution of the form $S = S_0 + S_1 e^{-r/r_0}$.  (\emph{b})~The
\emph{same} kinematic models, calculated using a WFPC2
H$\alpha$+[N~II] image as a more accurate map of the emission-line
surface brightness.  This figure demonstrates that many of the
irregularities seen in the rotation curve are not the result of
kinematic disturbances, but are simply due to the patchy
distribution of emission-line light across the slit.}
\end{figure}

\section{Direct Fitting of Emission-Line Profiles.}

The analysis technique described above, based on fitting the measured
velocity curves, only works well if the emission-line profiles are
close to Gaussian.  This is not always the case; for example, the
galaxy M84 (NGC 4374) has extremely complex line profiles near the
nucleus.  Bower et al.\ (1998) found that the line profiles in M84
show two kinematic components, a fast Keplerian component and a slowly
rotating one.  They fitted Keplerian disk models to the fast component
and derived a central mass of $(0.9-2.6)\times10^9$ $M_\odot$.  More
recently, Maciejewski \& Binney (2001) showed that the two kinematic
components actually both arise from the disk itself, because the slit
is wide enough to subtend different portions of the disk having a wide
range of velocities.  As a result, the position and velocity
information become badly entangled in the central regions, and it is
effectively impossible to measure a well-defined rotation curve for
the inner disk.

To circumvent this problem, we have adapted our code to fit the
emission profiles of the rotating disk \emph{directly} in pixel space.
This method makes use of all available information in the data,
including the high-velocity wings of the line profiles.  Preliminary
results for M84 are shown in Figure 2.  With $M_{\mathrm{BH}} = 10^9$
$M_\odot$, the model naturally accounts for much of the structure in
the line profiles: narrow cores and broad wings at the nucleus,
extreme asymmetries, and even double-peaked profiles at some
locations.  While the fits are by no means perfect, these preliminary
results demonstrate that direct profile fitting is a promising
technique for extracting black hole masses from data showing highly
asymmetric emission lines.

\begin{figure}
\plottwo{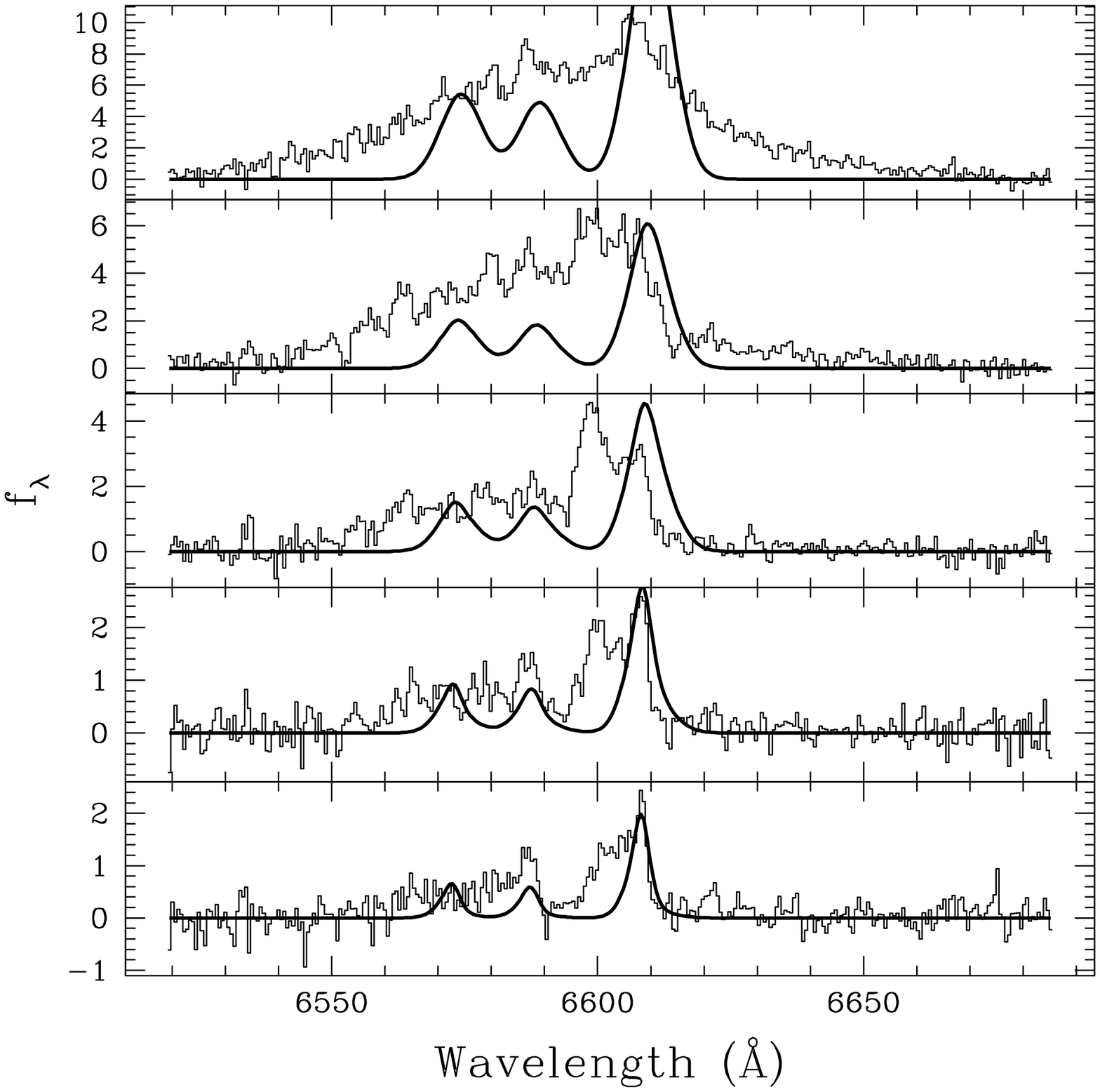}{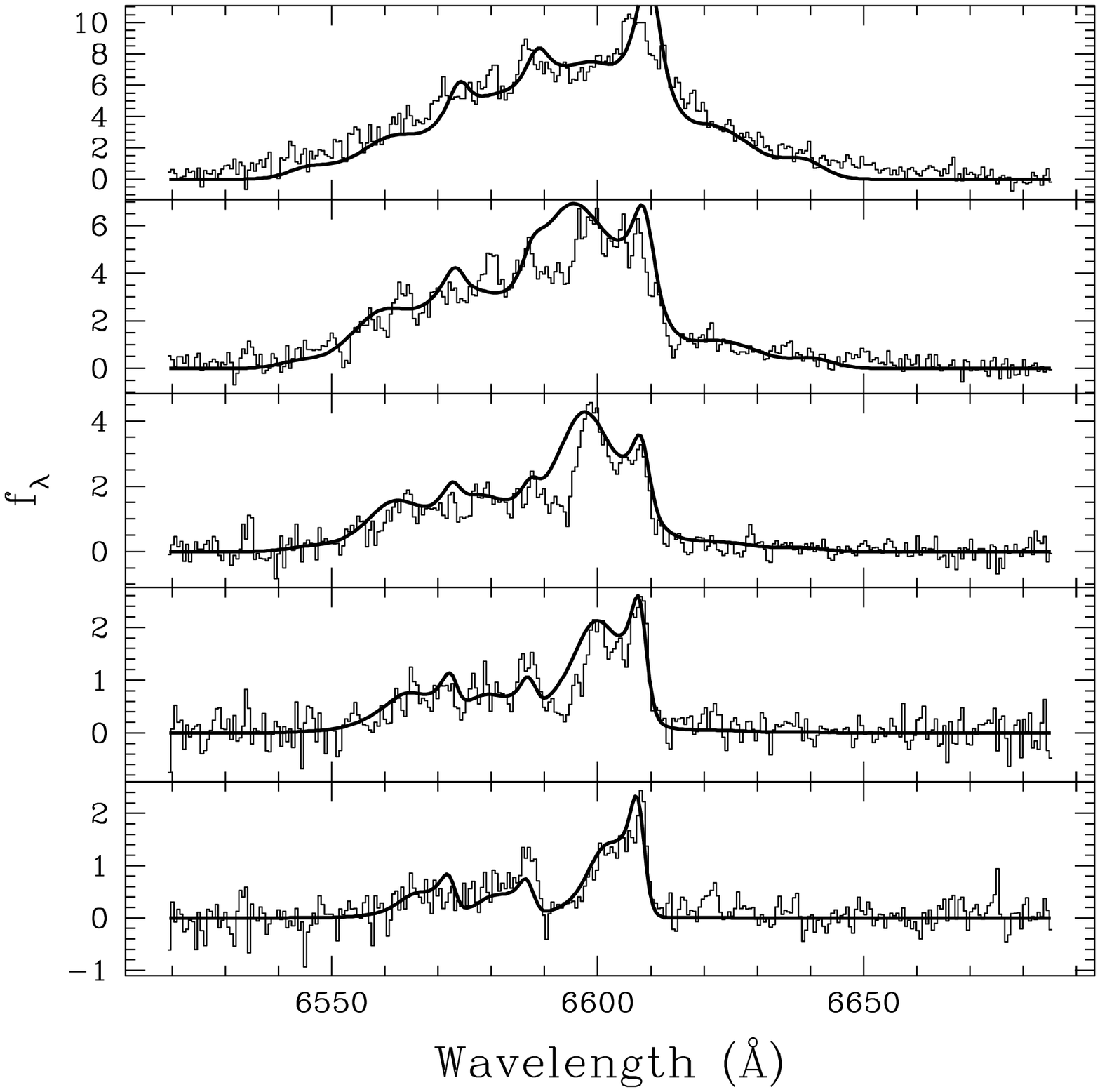}
\caption{Preliminary model results for M84.  Data and models are shown
for the central row of the CCD spectrum (top panel) and for four
off-nuclear rows. \emph{Left panel:} A trial model assuming no black
hole.  Without a black hole, the emission profiles are nearly Gaussian
and the model is unable to match the width and structure of the
observed profiles.  \emph{Right panel:} A model calculated for
$M_{\mathrm{BH}}=10^9$ $M_\odot$.}
\end{figure}

\end{document}